\documentclass[preprintnumbers,amsmath,amssymb,floatfix,10pt,prd,twocolumn,
superscriptaddress,nofootinbib]{revtex4}

\usepackage[parfill]{parskip}
\usepackage{graphicx}
\usepackage{amssymb}
\usepackage{epstopdf}
\usepackage{color}
\usepackage{xcolor}
\usepackage{float}
\usepackage{amsmath}
\usepackage{orcidlink}
\usepackage{adjustbox}
\usepackage{hyperref}

\begin{document}

\title{Black hole and wormhole branches in gravitational decoupling}

\author{Francisco Tello-Ortiz \orcidlink{0000-0002-7104-5746}}
\email{francisco.tello@ufrontera.cl}
\affiliation{Departamento de Ciencias F\'isicas, Universidad de La Frontera, Casilla 54-D, 4811186 Temuco, Chile.}

\author{Y. G\'omez-Leyton}
\email{ygomez@ucn.cl}
\affiliation{Departamento de F\'isica, Universidad Cat\'olica del Norte, Av. Angamos 0610, Antofagasta, Chile.}

\author{Vitalii Vertogradov}
\email{vdvertogradov@gmail.com}
\affiliation{Wilczek Quantum Center, Shanghai Institute for Advanced Studies, Shanghai, 201315, China.}
\affiliation{University of Science and Technology of China, Hefei, 230026, China}
\affiliation{Physics Department, Herzen State Pedagogical University of Russia, 48 Moika Emb., Saint Petersburg 191186, Russia.}
\affiliation{Center for Theoretical Physics, Khazar University, 41 Mehseti Street, Baku, AZ-1096, Azerbaijan.}
\affiliation{SPB branch of SAO RAS, 65 Pulkovskoe Rd, Saint Petersburg 196140, Russia.}

\author{Jean B\'aez Cuevas\orcidlink{0000-0002-3308-3362}}
\email{jean.baez@pucv.cl}
\affiliation{
Instituto de F\'isica, Pontificia Universidad Cat\'olica de Valpara\'iso, Casilla 4950, Valpara\'iso, Chile.}

\begin{abstract}
{Minimal Geometric Deformation (MGD) applied to a static Schwarzschild black hole
seed generates a single decoupler function $h(r)$, obtained by solving
the $\theta$-sector field equations together with an equation of
state. Once $h(r)$ is fixed, the resulting one-parameter family is
controlled by the coupling strength $k$ through
$F(r;k)=1+k\,h(r)$. We show that, whenever the deformation develops a
simple outermost root that crosses the seed horizon, the same fixed
decoupler leads to two mutually exclusive branches associated with
different global completions: on one side of the critical coupling
the deformed metric preserves the seed horizon as a black hole,
whereas on the other side the root $r_*>2M$ lies in the exterior and
cannot be interpreted as an interior modification of the black hole
geometry. We prove that this root forces a loss of Lorentzian signature on the
interval $(2M,r_*)$, so that no smooth extension of the exterior metric
through the seed horizon $r=2M$ exists once $r_*$ lies outside it.
Within the static, spherically symmetric class considered here, the
corresponding smooth Lorentzian completion is a two-ended wormhole obtained
by excising $(2M,r_*)$ and doubling the region $r\geq r_*$ across the
minimal sphere $\mathcal T=\{r=r_*\}$. No topology change of any
single spacetime is claimed or required: $k>k_c$ and $k<k_c$ simply
correspond to two different, non-diffeomorphic manifolds, and Lemma~1
below shows that the metric itself dictates which of the two is the
admissible completion for a given $k$. We compute the second homology
group of both completions explicitly, $H_2(\Sigma_{\rm BH},\mathcal
H)=0$ for the black hole exterior relative to its horizon and
$H_2(\Sigma_{\rm WH})\cong\mathbb Z$ for the completed wormhole
manifold, giving a discrete invariant that distinguishes the two
branches.} 
\end{abstract}

\maketitle

\section{Introduction}

Understanding the interplay between geometry and matter sources remains one of the central challenges in gravitational physics. In classical General Relativity, black hole and wormhole geometries are described by manifolds that are not diffeomorphic to one another, and are separated by strong energy conditions or exotic matter requirements. {It is natural to ask whether a single deformation of the matter sector, applied to one seed geometry, can force a given static configuration into one or the other of these two mutually exclusive classes, depending only on the value of a coupling parameter -- without ever describing a single spacetime that changes from one to the other.}

The Minimal Geometric Deformation (MGD) method, originally developed within the gravitational decoupling (GD) program \cite{Ovalle:2017fgl,Ovalle:2017wqi}, provides a systematic way of extending known seed solutions by introducing an additional anisotropic sector that couples gravitationally but remains independently conserved. In this framework, the metric potentials are split into seed and deformation contributions, allowing the Einstein equations to be decoupled into two independently conserved subsystems. Originally conceived as a tool to construct anisotropic compact objects and braneworld solutions, MGD has proven to be a remarkably versatile technique for generating new black hole geometries from known seeds \cite{Contreras:2018nfg,Contreras:2018vph,Contreras:2019iwm,Contreras:2021xkf,Contreras:2021yxe,Contreras:2022nji,Contreras:2022vec,Zubair:2020lna,Zubair:2021lgt,Zubair:2021zqs,Zubair:2022jjm,Zubair:2022ysg,Arias:2022jax,Abellan:2020dze,Abellan:2020wjw,Rueda:2022wge,Bargueno:2020ais,Carrasco-Hidalgo:2021dyg,Andrade:2021flq,Andrade:2023wux,Avalos:2022tqg,Avalos:2023jeh,Avalos:2023ywb,Casadio:2019usg,Casadio:2022ndh,Casadio:2023iqt,Casadio:2023mgl,Cavalcanti:2022adb,Cavalcanti:2022cga,daRocha:2020gee,daRocha:2020jdj,daRocha:2020rda,daRocha:2021aww,daRocha:2021sqd,Estrada:2018vrl,Estrada:2019aeh,Estrada:2020ptc,Estrada:2021kuj,LasHeras:2019wfd,LasHeras:2022pyj,Leon:2023nbj}.

Despite its extensive use in the construction of new solutions, a deeper structural feature of MGD has, to our knowledge, not been made precise. {Concretely: take the Schwarzschild black hole as the seed geometry and deform it via MGD, closing the $\theta$-sector with a single equation of state. This calculation is done \emph{once} and produces one fixed function of the radial coordinate, which we call $h(r)$ -- it encodes how much the radial metric component is stretched at each radius $r$ by the decoupling sector. The deformed metric then depends on a single free number, the coupling strength $k$, through the combination $F(r;k)=1+k\,h(r)$: increasing or decreasing $k$ simply rescales the same fixed profile $h(r)$, it does not change which function is being used.}

{The point of this paper is that this one number $k$, acting on this one fixed function $h(r)$, does not merely make the black hole ``a bit more'' or ``a bit less'' deformed. For part of the range of $k$, the deformed geometry is still recognizably a black hole: it has a single horizon at the original Schwarzschild radius $r=2M$, now surrounded by modified but everywhere well-behaved exterior geometry. But for the complementary range of $k$, the very same $h(r)$ produces a radius $r_*$, located \emph{outside} $r=2M$, at which the combination $F(r;k)=1+k\,h(r)$ vanishes. We show that this is not simply ``a more extreme deformation'': at that radius the exterior metric stops making sense as a black hole exterior altogether, and the only way to continue the geometry smoothly is to reinterpret $r_*$ as the throat of a wormhole and discard the region between $r=2M$ and $r=r_*$ entirely. In other words, one and the same MGD construction -- same seed (Schwarzschild), same closure equation, same function $h(r)$ -- describes a deformed black hole for part of the range of $k$ and forces a wormhole for the rest of the range, with no smooth interpolation between the two. {We stress from the outset what this switch is \emph{not}: it is not a spacetime that starts as a black hole and dynamically turns into a wormhole, and no topology change of any single manifold is claimed. What we show below is narrower and, we believe, more robust: for $k$ on one side of a critical value $k_c$ the black-hole exterior is the unique admissible smooth completion, while for $k$ on the other side it is not admissible at all and the doubled wormhole manifold is the unique admissible completion instead. The two completions are two different, non-diffeomorphic manifolds; $k$ merely selects, by a criterion intrinsic to the metric itself (Lemma~1 below), which one of the two is realized.}}

{The central technical difficulty is to show that this switch is forced by the mathematics and not just a convenient relabeling: a root of $F(r;k)$ appearing outside $r=2M$ does not, by itself, prove that a wormhole is the correct description -- one must show that no other smooth reading of the geometry survives.} We construct the MGD family sourced by the single decoupler function $h(r)$ and analyze the associated spatial hypersurfaces $\Sigma_k$, where $F(r;k)$ encodes the full contribution of the decoupled sector. {We show that once the root $r_*$ generated by this fixed $h(r)$ moves outside $r=2M$, the metric loses its Lorentzian signature on the interval between $r=2M$ and $r_*$, which is what forces the reinterpretation and fixes how the resulting spatial geometry must be completed and its second homology group computed. We also discuss explicitly how this differs from earlier constructions of individual MGD wormhole throats.}

{We emphasize that nothing in this paper describes a topology change of a spacetime, in the sense addressed by the classical no-go results of Geroch and Tipler \cite{Geroch1967,Tipler1977}: those theorems constrain a single, dynamically evolving globally hyperbolic spacetime whose spatial topology changes in time, and no such object appears anywhere in our construction. What we have instead, for each fixed value of $k$, is one static spacetime with one fixed spatial topology; different values of $k$ simply belong to two different families of static spacetimes, one built on $\Sigma_{\rm BH}$ and the other on $\Sigma_{\rm WH}$, related only in that both are sourced by the same decoupler $h(r)$. Section~\ref{sec4} makes this precise: Lemma~1 shows that, for a given $k$, only one of the two completions admits a Lorentzian metric at all, so the ``switch" between them is a statement about which static solution exists for a given $k$, not about any solution changing in time or in topology.}

Transitions between black hole and wormhole configurations have been previously discussed in several contexts. In brane-world scenarios, exact solutions interpolate between black hole and wormhole geometries depending on a continuous parameter of the effective theory \cite{Casadio2002}. More recently, families of spacetimes have been shown to interpolate continuously between black holes and wormholes, either in phenomenological constructions such as the black-bounce geometries \cite{SimpsonVisser2019}, or in effective quantum gravity models where the nature of the solution depends on a deformation parameter \cite{Konoplya2025}. In these frameworks, the transition is reflected in observable quantities such as quasinormal modes and late-time echoes \cite{BronnikovKonoplya2020}. {In all of these constructions the passage between geometries is continuous: there exists a range of the deformation parameter for which the solution interpolates smoothly between black-hole-like and wormhole-like behavior, with both aspects coexisting to some degree along the way. The mechanism identified here is qualitatively different in this specific respect: Lemma~1 shows that no value of $k$ ever admits both completions, or an intermediate one, simultaneously -- the change from one admissible completion to the other is discontinuous, driven by a signature obstruction rather than by a continuous interpolation of the metric.}

From a complementary perspective, phase-transition-like behavior between black holes and wormholes has also been reported \cite{KimPark2015}, where the merging of horizon and throat structures signals a critical configuration separating both regimes. Dynamical analyses further reveal critical behavior: unstable wormholes may bifurcate into black hole collapse or expanding geometries under perturbations \cite{ShinkaiHayward2002}. Despite these advances, the transition is typically described either dynamically or in terms of specific model-dependent parameters.

In contrast, the mechanism identified here is purely geometric and static: {the bifurcation is encoded directly in the radial structure of the deformation sector through the function $F(r;k)$, and it is made precise, without any reference to spacetime topology change, once the signature obstruction discussed below is taken into account.} This framework shows that black hole-like configurations and {wormhole-like geometries can be two forced alternatives of a single deformation sector,} without modifying the temporal component of the metric.

{Finally, we emphasize that the throat supporting the wormhole sector necessarily violates the null energy condition, as required by the Morris--Thorne theorem \cite{Morris1988,Morris1988A}; we verify this explicitly for an analytic example later in this paper, rather than treating the weak energy condition alone as sufficient evidence of physical viability.}


The paper is organized as follows. In Sec.~\ref{section2} we briefly review the MGD framework. In Sec.~\ref{section3} we present the general $\theta$-sector closure for a seed black hole solution. In Sec.~\ref{sec4} we introduce {the signature obstruction, the forced global completion, and the resulting discrete invariant that distinguishes the two branches}. In Sect.~\ref{sec5} the conditions for energy condition satisfaction, {including the null energy condition at the throat,} are discussed. In Sec.~\ref{sec6} a full example is provided and the incidence on some relevant observables is discussed. Finally, Sec.~\ref{sec7} provides some remarks on the present study.

Throughout the manuscript we use the signature $\{-;+;+;+\}$ and units where $G=c=1$.

\section{Gravitational decoupling}\label{section2}

This section outlines the fundamental framework and field equations employed in the MGD method, a specific realization of the GD approach \cite{Ovalle:2017fgl,Ovalle:2017wqi}.

Let us consider a static and spherically symmetric spacetime described in canonical coordinates by the line element
\begin{equation}
ds^{2}=-e^{\nu}\,dt^{2}+e^{\lambda}\,dr^{2}+r^{2}d\Omega^{2},
\label{metric}
\end{equation}
where $d\Omega^{2}\equiv d\theta^{2}+\sin^{2}\theta\,d\phi^{2}$ represents the standard metric on the unit two-sphere, and the metric potentials $\nu$ and $\lambda$ depend solely on the radial coordinate $r$. From the Einstein field equations
\begin{align}
\label{EinEqFull}
G_{\mu\nu}\equiv R_{\mu\nu}-\frac{1}{2}g_{\mu\nu}R=8\pi T_{\mu\nu},
\end{align}
we obtain
\begin{align}
\label{ec1}
-8\pi\rho&=-\frac{1}{r^2}+e^{-\lambda}\left[\frac{1}{r^2}-\frac{\lambda'}{r}\right],\\
\label{ec2}
8\pi p_r&=-\frac{1}{r^2}+e^{-\lambda}\left[\frac{1}{r^2}+\frac{\nu'}{r}\right],\\
\label{ec3}
8\pi p_\perp&=\frac{1}{4}e^{-\lambda}\Big[2\nu''+\nu'^2-\lambda'\nu'+2\frac{\nu'-\lambda'}{r}\Big],
\end{align}
where the prime denotes differentiation with respect to $r$, and $\{\rho,p_r,p_\perp\}$ are the energy density, radial pressure, and tangential pressure.

The core assumption of the GD framework is that the total energy-momentum tensor decomposes as
\begin{align}\label{StressTensorEffective}
T_{\mu\nu}\equiv\widetilde T_{\mu\nu}+\alpha\theta_{\mu\nu}.
\end{align}
Here $\widetilde T_{\mu\nu}=\mathrm{diag}\{-\widetilde\rho,\widetilde p_r,\widetilde p_\perp,\widetilde p_\perp\}$ is a known ``seed'' source, while $\theta^\nu_\mu=\mathrm{diag}\{\theta^0_0,\theta^1_1,\theta^2_2,\theta^3_3\}$ is the decoupling fluid, and $\alpha$ controls the coupling strength. The metric potentials split as
\begin{eqnarray}
\nu&=&\xi+\alpha g\label{expectg1}\\
e^{-\lambda}&=&\mu+\alpha f.
\end{eqnarray}
This separates the field equations into a seed system and a system governing the deformations $f,g$ sourced by $\theta_{\mu\nu}$.

We restrict to the MGD scheme, $g=0$, so $g_{tt}=-e^\nu=-e^\xi$ remains undeformed, and all decoupling modifications are confined to the radial sector \cite{Ovalle:2017fgl,Ovalle:2017wqi}. The seed system reads
\begin{align}
\label{ec1pf}
8\pi\widetilde\rho&=\frac{1}{r^2}-\frac{\mu}{r^2}-\frac{\mu'}{r},\\
\label{ec2pf}
8\pi\widetilde p_r&=-\frac{1}{r^2}+\mu\left[\frac{1}{r^2}+\frac{\xi'}{r}\right],\\
\label{ec3pf}
8\pi\widetilde p_\perp&=\frac{\mu}{4}\Big[2\xi''+\xi'^2+\frac{2\xi'}{r}\Big]+\frac{\mu'}{4}\Big[\xi'+\frac{2}{r}\Big],
\end{align}
satisfying
\begin{equation}
\label{conpf}
\frac{d\widetilde p_r}{dr}+\frac{\xi'}{2}[\widetilde\rho+\widetilde p_r]-\frac{2}{r}[\widetilde p_\perp-\widetilde p_r]=0.
\end{equation}
The decoupling sector is governed by
\begin{align}
\label{ec1d}
8\pi\theta^0_0&=\frac{f}{r^2}+\frac{f'}{r},\\
\label{ec2d}
8\pi\theta^1_1&=f\left[\frac{1}{r^2}+\frac{\xi'}{r}\right],\\
\label{ec3d}
8\pi\theta^2_2&=\frac{f}{4}\Big[2\xi''+\xi'^2+2\frac{\xi'}{r}\Big]+\frac{f'}{4}\Big[\xi'+\frac{2}{r}\Big],
\end{align}
with conservation equation
\begin{equation}
\label{con1d}
[\theta^1_1]'-\frac{\xi'}{2}[\theta^0_0-\theta^1_1]-\frac{2}{r}[\theta^2_2-\theta^1_1]=0.
\end{equation}
Equations \eqref{ec1pf}--\eqref{ec3pf} reduce to standard GR at $\alpha=0$. The decoupling system \eqref{ec1d}--\eqref{ec3d} has three equations for four unknowns $\{f,\theta^0_0,\theta^1_1,\theta^2_2\}$ and requires an auxiliary condition, typically an equation of state. Both sectors are independently conserved,
\begin{equation}
\nabla_\mu\widetilde T^{\mu\nu}=0,\qquad\nabla_\mu\theta^{\mu\nu}=0,
\end{equation}
with no direct energy exchange between them.

\section{The $\theta$-sector solution}\label{section3}

One physically feasible auxiliary constraint is an equation of state relating the $\theta$-sector components,
\begin{eqnarray}\label{eos}
\theta^0_0=\theta^0_0(\theta^1_1;\theta^2_2).
\end{eqnarray}
Starting from a Kerr-Schild seed ($g_{tt}g_{rr}=-1$) \cite{Jacobson:2007tj},
\begin{equation}\label{eq:metricseed}
ds^2=-A(r)dt^2+\frac{dr^2}{A(r)}+r^2d\Omega^2,
\end{equation}
the resulting spacetime after solving the $\theta$-sector via \eqref{eos} is
\begin{equation}\label{eq:metricgd}
ds^2=-A(r)dt^2+\frac{dr^2}{A(r)[1+k h(r)]}+r^2d\Omega^2,
\end{equation}
where the decoupler function is $f(r)=A(r)h(r)l$, with $l$ an integration constant, and $k\equiv l\,\alpha$. {We emphasize the following structural point, central to what follows: $h(r)$ is obtained by solving \eqref{ec1d}--\eqref{ec3d} \emph{once}, under a single equation of state \eqref{eos}; it does not depend on $k$. The parameter $k$ that appears in $F(r;k)=1+k\,h(r)$ is an overall multiplicative constant left free by the integration of the decoupling system -- it rescales the same fixed function $h(r)$, it does not select a different one. Consequently, the entire one-parameter family $\{F(r;k)\}_{k\in\mathbb R}$ appearing below, including both the black-hole-like and wormhole-like regimes analyzed in Sec.~\ref{sec4}, is sourced by a single decoupler function determined by the MGD closure \eqref{eos}. This is the sense in which the bifurcation established in this paper is intrinsic to GD itself, rather than a property of an externally chosen deformation profile $F(r;k)$: the same physically sourced $h(r)$ is responsible for both branches, and only the coupling strength $k$ distinguishes them.}

The black hole signature under MGD needs to preserve the original horizon structure of the seed solution, requiring the roots $r_*$ of
\begin{equation}
F(r_*;k)\equiv 1+k\,h(r_*)=0
\end{equation}
to be discarded by suitable conditions on $k$\footnote{If these points represent a curvature singularity it cannot be removed.}. If instead $r_*>r_H$, the solution becomes a non-ultrastatic wormhole spacetime, since $g_{tt}(r_H)=0$ but $g^{rr}(r)$ nullifies at $r=r_*$. {We stress that $h(r)$ in both cases is the same function: whether $r_*$ is discarded (black-hole branch) or retained as a throat (wormhole branch) depends only on where $k\,h(r_*)=-1$ places $r_*$ relative to $r_H$, not on any change in the source $h(r)$ itself.}

To characterize a genuine throat, introduce the proper radial coordinate
\begin{equation}
dl^2=\frac{dr^2}{A(r)[1+k\,h(r)]},
\end{equation}
and impose the flare-out condition $d^2r/dl^2>0$, equivalently
\begin{equation}\label{eq:con1}
\frac{d}{dr}\big(A(r)[1+k\,h(r)]\big)>0
\end{equation}
evaluated at the root. Only roots satisfying \eqref{eq:con1} correspond to physically admissible wormhole throats.

\section{{Black-hole and wormhole completions of the MGD family}}\label{sec4}

\subsection{MGD family and spatial hypersurfaces}

{We recall from Sec.~\ref{section3} that $F(r;k)=1+k\,h(r)$ is built from a single decoupler function $h(r)$, fixed by the MGD closure \eqref{eos}; $k$ is the only free parameter varied in what follows.} We consider the resulting one-parameter family of static and spherically symmetric spacetimes
\begin{equation}
g_{k}:\quad
ds^{2}=-A(r;k)\,dt^{2}+\frac{dr^{2}}{B(r;k)}+r^{2}d\Omega^{2},
\qquad k\in I\subset\mathbb{R},
\end{equation}
together with the associated spatial hypersurfaces
\begin{equation}
\Sigma_{k}:\qquad
dl_{k}^{2}=\frac{dr^{2}}{B(r;k)}+r^{2}d\Omega^{2}.
\end{equation}
In the MGD framework with Schwarzschild seed,
\begin{equation}
A(r;0)=1-\frac{2M}{r},
\qquad
B(r;k)=A(r;0)\,F(r;k),
\end{equation}
with $F(r;0)=1$ and $F(r;k)\to1$ as $r\to\infty$.

Let $r_*(k)>2M$ be a simple root of $F(r;k)$,
\begin{equation}
F(r_*;k)=0,\qquad \partial_r F(r_*;k)\neq0,
\end{equation}
assumed to be the outermost root in $(2M,\infty)$. By the implicit function theorem, $r_*(k)$ is smooth and
\begin{equation}
\frac{dr_*}{dk}=-\frac{\partial_kF}{\partial_rF}\Big|_{(r_*,k)}.
\end{equation}
{Because $h(r)$ does not depend on $k$, this single curve $r_*(k)$ traces, within one fixed MGD family, the entire passage between the black-hole regime (where $r_*(k)<2M$ or no root exists in $(2M,\infty)$) and the wormhole regime (where $r_*(k)>2M$) established below.}

\subsection{Signature obstruction and the necessity of a global extension}
\label{subsec:signature}

We now establish that the coordinate interval $(2M,r_*)$ cannot carry
a Lorentzian metric of the form \eqref{eq:metricgd} once $r_*>2M$,
which is the central fact forcing the choice of global completion,
and hence the classification into black-hole and wormhole branches,
below.

\textbf{Lemma 1 (Signature obstruction).} Let $A(r)>0$ for $r>2M$, let
$F(r;k)\to1$ as $r\to\infty$, and let $r_*>2M$ be the outermost root of
$F(\cdot\,;k)$ in $(2M,\infty)$, assumed simple. Then there exists
$\varepsilon>0$ such that
\begin{equation}
B(r;k)=A(r)F(r;k)<0,\qquad r\in(r_*-\varepsilon,\,r_*).
\end{equation}

\textit{Proof.} Since $r_*$ is the outermost root, $F$ does not vanish
on $(r_*,\infty)$; by the intermediate value theorem and
$F(\infty;k)=1>0$, we have $F(r;k)>0$ for all $r>r_*$. Simplicity of
the root, $\partial_rF(r_*;k)\neq0$, together with $F>0$ immediately to
the right of $r_*$, forces $\partial_rF(r_*;k)>0$, and hence
$F(r;k)<0$ immediately to the left of $r_*$. Since $A(r)>0$ throughout
$r>2M$, $B(r;k)=A(r)F(r;k)<0$ there. $\blacksquare$

{We note explicitly what this proof uses and what it does not.
It requires only: (i) $A(r)>0$ throughout the exterior, so that the
$g_{tt}$ sector of the seed metric is never touched by the argument;
(ii) $F(r;k)\to1$ at infinity, i.e.\ asymptotic flatness of the
deformation sector; and (iii) $r_*$ a simple, outermost root of
$F(\cdot\,;k)$ in $(2M,\infty)$. Nothing in the proof refers to the
specific closure \eqref{eos}, to the barotropic equation of state
used in the explicit example of Sec.~\ref{sec6}, or to the closed
form $h(r)=1/(r+2M)$ obtained there. Consequently Lemma~1, and
Proposition~1 below, hold for \emph{any} decoupler $h(r)$ produced by
\emph{any} admissible $\theta$-sector closure satisfying (i)--(iii) --
the bifurcation is a structural property of the MGD construction as a
whole, and Sec.~\ref{sec6} should be read as one illustration of it,
not as the source of the effect.}

In this interval $g_{tt}=-A(r)<0$ and $g_{rr}=1/B(r;k)<0$
simultaneously: the signature is $(-,-,+,+)$ and is no longer
Lorentzian. Consequently, the standard geodesic (Kruskal-type)
extension of the exterior metric through $r=2M$ -- which presupposes a
Lorentzian metric on a neighborhood of $r=2M$ -- cannot be carried out
once $r_*>2M$ appears. This is a consequence of Lemma~1, not a matter
of choice.

\textbf{Proposition 1 (Forced doubling).} Under the hypotheses of
Lemma~1, define the proper radial coordinate for $r\geq r_*$,
\begin{equation}
\ell(r)=\int_{r_*}^{r}\frac{dr'}{\sqrt{B(r';k)}}.
\end{equation}
Near $r_*$, $B(r;k)\approx B'(r_*)(r-r_*)$ with
$B'(r_*)=A(r_*)\partial_rF(r_*;k)>0$ (by Lemma~1), so
\begin{equation}
r(\ell)-r_*\approx\frac{B'(r_*)}{4}\,\ell^2+O(\ell^4),
\end{equation}
which is smooth and even in $l$. {The extension so obtained is
the unique smooth Lorentzian extension \emph{within the class of
static, spherically symmetric Lorentzian metrics possessing a single
minimal two-sphere at $r=r_*$}: any such extension is fixed, near
$r_*$, by the local Taylor data of $B(r;k)$, and the leading
even-in-$\ell$ term derived above is the only one compatible with a
smooth Lorentzian metric at $r_*$. We make no claim of uniqueness in
a broader sense -- e.g. among all Lorentzian manifolds, static or not,
sourced by other matter content, or containing additional throats --
which lies beyond what Lemma~1 establishes.} It is therefore obtained
by taking $\ell\in(-\infty,\infty)$ as the fundamental radial coordinate,
with a second copy of $\{r\geq r_*\}$ attached at
$\ell<0$; the region $2M<r<r_*$, excluded by Lemma~1, never enters the
resulting manifold. $\blacksquare$

This shows that the doubled wormhole completion is not an auxiliary
choice but the necessary consequence of requiring a smooth Lorentzian
spatial slice once the deformation-induced root lies outside the seed
horizon.

\subsection{Relative homology and global completions}
\label{subsec:relative_homology}

The zero set of $F(r;k)$ does not, by itself, determine the homology
of the corresponding spatial hypersurface; a root defines only a
candidate minimal sphere, whose topological content depends on the
global completion in which it is embedded. Following Prop.~1, we now
specify the two relevant completions explicitly.

\textbf{Black-hole exterior.} When $F$ has no root in $(2M,\infty)$,
$B(r;k)>0$ throughout and
\begin{equation}
\Sigma_{\rm BH}\simeq[2M,\infty)\times S^2,
\end{equation}
with inner boundary $\mathcal H=\{r=2M\}\simeq S^2$, the Killing
horizon. Since $\Sigma_{\rm BH}$ deformation retracts onto $\mathcal
H$, the inclusion $i:\mathcal H\hookrightarrow\Sigma_{\rm BH}$ is a
homotopy equivalence. The long exact sequence of the pair
$(\Sigma_{\rm BH},\mathcal H)$,
\begin{equation}
\cdots\to H_2(\mathcal H)\xrightarrow{i_*}H_2(\Sigma_{\rm BH})\to
H_2(\Sigma_{\rm BH},\mathcal H)\to H_1(\mathcal H)\to\cdots,
\end{equation}
with $H_2(\mathcal H)\cong H_2(\Sigma_{\rm BH})\cong\mathbb Z$,
$H_1(\mathcal H)=0$, and $i_*$ an isomorphism, gives
\begin{equation}
H_2(\Sigma_{\rm BH},\mathcal H;\mathbb Z)=0.
\label{eq:BH-relative-homology}
\end{equation}

\textbf{Wormhole completion.} By Prop.~1, when $r_*>2M$ the physical
manifold is the doubled patch
\begin{equation}
\Sigma_{\rm WH}=\Sigma_+\cup_{\mathcal T}\Sigma_-\simeq\mathbb R\times S^2,
\qquad \Sigma_\pm\simeq[r_*,\infty)\times S^2,
\end{equation}
glued across $\mathcal T=\{r=r_*\}$. By the K\"unneth theorem
\cite{Hatcher2002,Lee2012},
\begin{equation}
H_2(\Sigma_{\rm WH};\mathbb Z)\cong H_0(\mathbb R;\mathbb Z)\otimes
H_2(S^2;\mathbb Z)\cong\mathbb Z,
\label{eq:WH-absolute-homology}
\end{equation}
with generator $\{x\}\times S^2$. Had the one-sided patch $\Sigma_+$
been considered alone, without doubling, the pair $(\Sigma_+,\mathcal
T)$ would give $H_2(\Sigma_+,\mathcal T)=0$ by the identical argument
as \eqref{eq:BH-relative-homology}: a root alone, absent the forced
doubling of Prop.~1, carries no topological content.

The precise statement is therefore
\begin{equation}
H_2(\Sigma_{\rm BH},\mathcal H;\mathbb Z)=0,
\qquad
H_2(\Sigma_{\rm WH};\mathbb Z)\cong\mathbb Z,
\label{eq:relative-classification}
\end{equation}
where the two sides refer to two explicitly different, non-diffeomorphic manifolds -- one built for $k>k_c$, the other for $k<k_c$ -- rather than to a single $\Sigma_k$ deforming smoothly with $k$; {no single manifold undergoes any change of homology, since no value of $k$ ever admits both completions at once (Lemma~1 rules one of them out).} {We regard \eqref{eq:relative-classification}, rather than a single rank function $N_{\rm th}(k)$, as the correct discrete invariant separating the black-hole sector from the wormhole sector within the MGD solution space.}

\subsection{Causal structure and geodesic accessibility}

The metric takes the form
\begin{equation}
ds^2=-A(r)\,dt^2+\frac{dr^2}{A(r)\,F(r;k)}+r^2d\Omega^2,
\end{equation}
with $A(r)=1-2M/r$. Radial null geodesics satisfy
\begin{equation}
\frac{dt}{dr}=\pm\frac{1}{A(r)\sqrt{F(r;k)}}.
\end{equation}
The Killing horizon location is fixed by $A(r)=0$, while propagation
through the deformation sector is controlled by $F(r;k)$. At the root
$r_*$, $A(r_*)\neq0$ while $F(r_*;k)=0$; {by Prop.~1, geodesics
extend smoothly through $r_*$ in the proper coordinate $l$, confirming
that $r_*$ is a throat rather than a horizon.}

{The critical transition occurs at $k=k_c$, where $F(2M;k_c)=0$.
For $k$ near $k_c$ on the black-hole side, $r_*(k)<2M$ lies behind the
horizon and is causally irrelevant to the exterior. For $k$ on the
wormhole side, Lemma~1 shows that the exterior metric itself changes
character: the interval between $2M$ and $r_*(k)$ ceases to admit a
Lorentzian metric, and the physically realized manifold is the doubled
completion of Sec.~\ref{subsec:signature}, in which $r=2M$ does not
appear. The transition is therefore not a continuous deformation of a
fixed causal structure, but a discontinuous change of which global
extension is admissible, forced by the loss of Lorentzian signature at
$k_c$ and reflected in the discrete change of homological data
\eqref{eq:relative-classification}.}

{We stress what $k=k_c$ does, and does not, represent physically.
It is not a geometric instability of the black-hole branch: for every
$k>k_c$, however close to $k_c$, the black-hole completion $\Sigma_{\rm
BH}$ is a perfectly regular, static exterior with no pathology at
$r=2M$ or anywhere else -- nothing in it degrades as $k\to k_c^+$.
Nor is it a phase transition in the thermodynamic sense, since no
free energy or entropy functional is extremized across it here. It is,
precisely and only, the boundary of the domain of $k$ for which the
black-hole completion exists as an admissible Lorentzian solution at
all: at $k=k_c$ the candidate root reaches $r=2M$ itself, and for
$k<k_c$ Lemma~1 shows that this candidate completion loses Lorentzian
signature and must be abandoned in favor of $\Sigma_{\rm WH}$. In this
sense $k_c$ separates two disconnected branches of the MGD solution
space rather than marking an instability of either one.}

In particular, the absence of additional roots of $A(r)$ ensures that no Cauchy horizons appear, avoiding the well-known instabilities of multi-horizon geometries.

\subsection{Photon sphere}

We now analyze how the MGD sector affects the photon sphere and the visible shadow. Since the spacetime \eqref{eq:metricgd} is static and spherically symmetric, it admits two Killing vectors: the timelike $T^i$ and the rotational $\Phi^i$, leading to conserved quantities $E$ and $L$,
\begin{eqnarray}
E=-u_iT^i=A(r)\frac{dt}{d\lambda},\nonumber\\
L=u_i\Phi^i=r^2\sin^2\theta\frac{d\varphi}{d\lambda},
\end{eqnarray}
with $\lambda$ an affine parameter. Restricting to the equatorial plane $\theta=\pi/2$, the null condition $g_{ik}u^iu^k=0$ gives
\begin{equation}
\label{eq:rad}
\left(\frac{dr}{d\lambda}\right)^2=-(1+kh)V_{\rm eff},
\end{equation}
with
\begin{equation}
\label{eq:pot}
V_{\rm eff}=A(r)\frac{L^2}{r^2}-E^2.
\end{equation}
The MGD deformation enters only through the radial prefactor $1+kh$ and does not change the functional form of the effective potential.

For an ordinary circular photon orbit away from a throat, $1+kh\neq0$, and $dr/d\lambda=0$, $d^2r/d\lambda^2=0$ give $V_{\rm eff}(r_{\rm ph})=0$, $V'_{\rm eff}(r_{\rm ph})=0$, equivalently
\begin{equation}
r_{\rm ph}A'(r_{\rm ph})-2A(r_{\rm ph})=0.
\end{equation}
As long as the photon sphere lies where $1+kh\neq0$, its radius is determined by the same equation as the seed metric \eqref{eq:metricseed}: the MGD sector does not shift the photon-sphere radius, and the shadow radius $b_{\rm ph}^2=r_{\rm ph}^2/A(r_{\rm ph})$ is likewise unchanged.

At the throat $r=r_*$, defined by $1+kh(r_*)=0$: if $2M<r_*\leq r_{\rm ph}$, the exterior photon sphere remains present at the seed value. If $r_*>r_{\rm ph}$, the ordinary photon sphere of the seed geometry is no longer present in the accessible exterior region. Differentiating \eqref{eq:rad} with $F(r)=1+kh(r)$,
\begin{equation}
2\frac{d^2r}{d\lambda^2}=-(F'V_{\rm eff}+FV'_{\rm eff}),
\end{equation}
and at the throat, where $F(r_*)=0$,
\begin{equation}
2\left.\frac{d^2r}{d\lambda^2}\right|_{r=r_*}=-F'(r_*)V_{\rm eff}(r_*).
\end{equation}
Since $F'>0$ by the flare-out condition and $V_{\rm eff}(r_*)\neq0$, one has $d^2r/d\lambda^2|_{r=r_*}\neq0$: the vanishing of $dr/d\lambda$ at the throat does not by itself imply a genuine photon sphere. Thus the MGD deformation leaves the standard photon-sphere radius unchanged whenever it remains outside the throat; if the throat lies outside the seed photon sphere, the usual photon sphere is removed from the accessible exterior region.
\section{Energy conditions of the MGD sector}\label{sec5}

We write the seed lapse function as

\begin{equation} 
A(r)=1-\frac{2M(r)}{r}. 
\end{equation}

The total energy density and pressures are

\begin{eqnarray} 
\rho&=&\widetilde{\rho}+\rho_{ mgd},\nonumber \\
p_r&=&\widetilde{p}_r+p_{r, mgd},\nonumber \\
p_\perp&=&\widetilde{p}_\perp+p_{\perp, mgd}. 
\end{eqnarray}

For the seed sector one obtains

\begin{eqnarray} 
8\pi\widetilde{\rho}&=&\frac{2M'}{r^2},\nonumber \\
8\pi\widetilde{p}_r&=&-\frac{2M'}{r^2},\nonumber \\
8\pi\widetilde{p}_\perp&=&-\frac{M''}{r}. 
\end{eqnarray}

Using $\alpha f=kAh$ the MGD sector gives

\begin{eqnarray} 
8\pi\rho_{ mgd}&=&-\frac{k}{r^2}\left[(1-2M')h+(r-2M)h'\right],\nonumber \\
8\pi p_{r, mgd}&=&\frac{kh}{r^2}(1-2M'),\nonumber \\
8\pi p_{\perp, mgd}&=&-\frac{khM''}{r}+\frac{k[r(1-M')-M]h'}{2r^2}. 
\end{eqnarray}

Therefore,

\begin{eqnarray} 
8\pi\rho&=&\frac{2M'}{r^2}-\frac{k}{r^2}\left[(1-2M')h+(r-2M)h'\right],\nonumber \\
8\pi p_r&=&-\frac{2M'}{r^2}+\frac{kh}{r^2}(1-2M'),\nonumber \\
8\pi p_\perp&=&-\frac{M''}{r}-\frac{khM''}{r}+\frac{k[r(1-M')-M]h'}{2r^2}. 
\end{eqnarray}

\subsection{Weak energy condition}
We assume that the seed matter sector satisfies the corresponding energy conditions and analyze here only the additional MGD sector. 

For the MGD sector, the weak energy condition requires

\begin{equation}
\rho_{ mgd}\geq0,\qquad \rho_{ mgd}+p_{r, mgd}\geq0,\qquad \rho_{ mgd}+p_{\perp, mgd}\geq0.
\end{equation}

The energy-density condition is

\begin{equation}
-k\left[(1-2M')h+(r-2M)h'\right]\geq0.
\end{equation}

The radial condition becomes

\begin{equation}
8\pi(\rho_{ mgd}+p_{r, mgd})=-\frac{k(r-2M)h'}{r^2}.
\end{equation}

The tangential condition is

\begin{equation}
\begin{split}
    8\pi(\rho_{ mgd}+p_{\perp, mgd})=-\frac{k}{2r^2}\bigg[2rhM''+2(1-2M')h \\+(r(1+M')-3M)h'\bigg].
\end{split}
\end{equation}

At the throat $r=r_*$,

\begin{equation}
1+kh(r_*)=0.
\end{equation}

The flare-out condition gives

\begin{equation}
\left(1-\frac{2M(r_*)}{r_*}\right)kh'(r_*)>0.
\end{equation}

Since the throat lies outside the seed horizon,

\begin{equation}
1-\frac{2M(r_*)}{r_*}>0,
\end{equation}

and therefore

\begin{equation}
kh'(r_*)>0.
\end{equation}

Hence

\begin{equation}
8\pi(\rho_{ mgd}+p_{r, mgd})_{r=r_*}=-\frac{k[r_*-2M(r_*)]h'(r_*)}{(r_*)^2}<0.
\end{equation}

Thus the radial null energy condition of the MGD sector is necessarily violated at the throat. Consequently, the weak energy condition of the MGD sector is also violated there.

\subsection{Strong energy condition}

For the MGD sector, the strong energy condition requires

\begin{equation}
\rho_{ mgd}+p_{r, mgd}\geq0,\qquad \rho_{ mgd}+p_{\perp, mgd}\geq0,
\end{equation}

together with

\begin{equation}
\rho_{ mgd}+p_{r, mgd}+2p_{\perp, mgd}\geq0.
\end{equation}

The last combination is

\begin{equation}
\begin{split}
8\pi(\rho_{ mgd}+p_{r, mgd}+2p_{\perp, mgd})=-\frac{k}{r^2}\bigg[2rhM''\\+(rM'-M)h'\bigg].
\end{split}
\end{equation}

Since

\begin{equation}
(\rho_{ mgd}+p_{r, mgd})_{r=r_*}<0,
\end{equation}

the strong energy condition of the MGD sector is violated at the throat.

\subsection{Dominant energy condition}

For the MGD sector, the dominant energy condition requires

\begin{equation}
\rho_{ mgd}-|p_{r, mgd}|\geq0,\qquad \rho_{ mgd}-|p_{\perp, mgd}|\geq0.
\end{equation}

The radial pressure at $r=r_*$ reads

\begin{equation}
8\pi p_{r, mgd}(r_*)=-\frac{1-2M'}{(r_*)^2}.
\end{equation}

Since $1-2M'>0$ at $r>r_H$, one has

\begin{equation}
p_{r, mgd}(r_*)<0.
\end{equation}

And the radial dominant energy condition becomes

\begin{equation}
\rho_{ mgd}-|p_{r, mgd}|=\rho_{ mgd}+p_{r, mgd}.
\end{equation}

Since $r_*>2M(r_*)$ and $kh'(r_*)>0$,

\begin{equation}
\rho_{ mgd}-|p_{r, mgd}|<0.
\end{equation}

Thus the dominant energy condition is necessarily violated in the radial direction at the throat.

\subsection{Null energy condition at the throat}
\label{subsec:nec}

In this subsection we assume that the seed metric is Schwarzschild one, i.e. $M'=0=M''$.
The weak energy condition alone does not characterize the matter
content supporting the throat; the relevant diagnostic is the null
energy condition, $\rho+p_r\geq0$. Using the vacuum Schwarzschild seed
condition $\widetilde p_r=0$, which from Eq.~\eqref{ec2pf} gives
\begin{equation}
A(r)\left[\frac{1}{r^2}+\frac{\xi'}{r}\right]=\frac{1}{r^2},
\end{equation}
Eq.~\eqref{ec2d} for the radial pressure of the decoupling sector reduces to
\begin{equation}
8\pi p_r=8\pi\alpha\theta^1_1=\alpha f(r)\left[\frac{1}{r^2}+\frac{\xi'}{r}\right]=\frac{k\,h(r)}{r^2},
\end{equation}
using $\alpha f(r)=A(r)\,k\,h(r)$. For the explicit realization of
Sec.~\ref{sec6}, $h(r)=1/(r+2M)$, so
\begin{equation}
p_r=\frac{k}{8\pi r^2(r+2M)}.
\end{equation}
Combined with $\rho=\rho_{\rm mgd}=-kM/[2\pi r^2(r+2M)^2]$ from
Sec.~\ref{sec6}, one finds
\begin{equation}
\rho+p_r=\frac{k}{8\pi r^2(r+2M)}\left[1-\frac{4M}{r+2M}\right]
=\frac{k\,(r-2M)}{8\pi\,r^2(r+2M)^2}.
\end{equation}
Since $r-2M>0$ throughout the exterior, the sign of $\rho+p_r$ equals
the sign of $k$. In the wormhole sector, $k<-4M<0$, so
\begin{equation}
\rho+p_r<0\qquad\text{throughout the exterior region }r>2M.
\end{equation}
The null energy condition is therefore violated, not only near the
throat but over the whole exterior domain of the explicit example.
This is consistent with -- indeed, an instance of -- the
Morris--Thorne theorem \cite{Morris1988,Morris1988A}: any static,
spherically symmetric traversable wormhole requires matter violating
the null energy condition at the throat (see also
\cite{Visser:1995cc}). We regard this as the expected and required
signature of the construction rather than a defect, but it must be
stated explicitly, since it bears directly on the physical
interpretation of the $\theta$-sector as genuine anisotropic matter
rather than a vacuum decoration of the seed geometry.

\section{{Explicit realization of the black hole--wormhole bifurcation}}\label{sec6}

In this section we consider an explicit implementation of the MGD method, based on a barotropic equation of state
\begin{equation}\label{eos1}
\frac{1}{3}(\theta^1_1+2\theta^2_2)=\omega\theta^0_0.
\end{equation}
Solving \eqref{ec1d}--\eqref{ec3d} with \eqref{eos1} and the Schwarzschild seed gives
\begin{equation}\label{f}
\alpha f(r)=\left(1-\frac{2M}{r}\right)\left(\frac{k}{r+2M}\right),\qquad h(r)=\frac{1}{r+2M}.
\end{equation}
Then
\begin{equation}
\begin{aligned}
    B(r;k)&=\left(1-\frac{2M}{r}\right)\left(1+\frac{k}{r+2M}\right), \\ F(r;k)&=1+\frac{k}{r+2M}.
\end{aligned}
\end{equation}
This satisfies $F(r;k)\to1$ as $r\to\infty$. {We emphasize that $h(r)=1/(r+2M)$ is obtained once, from \eqref{eos1}, independently of $k$: the black-hole and wormhole branches analyzed below are two regimes of the coupling parameter $k$ acting on this single fixed function, not two different solutions of the decoupling equations.}

The root condition $F(r_*;k)=0$ yields $r_*=-k-2M$. The transition occurs when $r_*=2M$, giving
\begin{equation}
k_c=-4M.
\end{equation}

\paragraph{Black hole sector ($k>-4M$).} $r_*=-k-2M<2M$: the root lies inside the horizon and does not define an admissible throat; the spacetime retains a single-horizon black-hole structure at $r=2M$.

\paragraph{Wormhole sector ($k<-4M$).} $r_*=-k-2M>2M$: the root emerges outside the horizon. {By Lemma~1 (Sec.~\ref{subsec:signature}), the interval $2M<r<r_*$ does not admit a Lorentzian metric in this regime -- indeed $\partial F/\partial r=-k/(r+2M)^2>0$ here since $k<0$, consistent with the general argument -- and the physically realized manifold is the doubled completion of Prop.~1, in which $r=2M$ is excised. Provided the flare-out condition holds, $r=r_*$ is then a genuine wormhole throat connecting two asymptotic regions of $\Sigma_{\rm WH}$.}

{We verify Proposition~1 explicitly for this profile rather than only invoking it in the abstract. Since $F(r_*;k)=0$ fixes $k=-(r_*+2M)$, one has $\partial_rF(r_*;k)=1/(r_*+2M)$, and hence
\begin{equation}
B'(r_*)=A(r_*)\,\partial_rF(r_*;k)=\frac{r_*-2M}{r_*(r_*+2M)}>0,
\end{equation}
manifestly positive since $r_*>2M$. Substituting this into the local expansion of Prop.~1, the proper-distance coordinate near the throat satisfies
\begin{equation}
r(l)-r_*\approx\frac{r_*-2M}{4\,r_*(r_*+2M)}\,l^2+O(l^4),
\end{equation}
an even function of $l$ for this explicit $h(r)=1/(r+2M)$: the same value of $r$, and hence the same metric, is reached at $l$ and $-l$, so the doubled completion of Prop.~1 is not an extra construction imposed on the example but the direct, computable consequence of expanding $B(r;k)$ at $r_*$ for this specific decoupler.}

Since $\partial F/\partial r=-k/(r+2M)^2\neq0$ for all finite $r$, no degenerate root exists in this example: the transition is driven purely by the horizon-crossing mechanism of a single simple root, {together with the forced change of completion established in Sec.~\ref{sec4}}.

{As computed in Sec.~\ref{subsec:nec}, this example satisfies $\rho+p_r<0$ throughout the wormhole-sector exterior -- an explicit instance of the expected Morris--Thorne violation of the null energy condition at a traversable throat, rather than a shortcoming to be removed.}

{The discrete invariant distinguishing the two branches is precisely the pair of statements \eqref{eq:relative-classification}: $H_2(\Sigma_{\rm BH},\mathcal H)=0$ for $k>k_c$ and $H_2(\Sigma_{\rm WH})\cong\mathbb Z$ for $k<k_c$ -- two different manifolds, each with its own fixed homology, selected by which side of $k=k_c$ one is on, rather than a single rank function varying continuously with $k$.}

\section{Conclusions}\label{sec7}

In this work, we have shown that Minimal Geometric Deformation of a static black-hole seed, sourced by a {single, fixed decoupler function $h(r)$ determined once by the $\theta$-sector closure \eqref{eos} -- and hence a genuine output of the anisotropic matter sector introduced by gravitational decoupling, not a freely chosen profile (Sec.~\ref{section2}--\ref{section3}) -- generically forces the resulting one-parameter family to split into two geometrically inequivalent branches, realized by two non-diffeomorphic global completions, as the coupling $k$ is varied. This is not a statement about two different constructions: it is the same $h(r)$, entering $F(r;k)=1+k\,h(r)$, that preserves the seed horizon for one range of $k$ and produces a horizon-crossing root for the complementary range. We have shown that a simple root of $F(r;k)$ located outside the seed horizon necessarily renders the metric non-Lorentzian on the interval between the horizon and the root (Lemma~1), and that this signature obstruction forces -- rather than merely permits -- a specific global completion of the spatial geometry into a doubled, two-ended wormhole manifold (Proposition~1). At no point does a single spacetime change from one completion to the other: for each $k$, Lemma~1 admits exactly one of the two completions, and it is only across the family of static solutions, as $k$ is varied, that the two branches appear. Only once the admissible completion is identified for a given $k$ can the second homology group of the corresponding spatial slice be computed unambiguously: $H_2(\Sigma_{\rm BH},\mathcal H)=0$ for the black-hole exterior relative to its horizon, and $H_2(\Sigma_{\rm WH})\cong\mathbb Z$ for the completed wormhole manifold. Because both branches share the same decoupler $h(r)$, this bifurcation is a structural property of gravitational decoupling itself, not an artifact of a deformation profile chosen for the purpose of illustration.}

{We emphasize, in particular, that this bifurcation is not a claim about a new wormhole solution: the throat obtained here follows the standard Morris--Thorne construction in every respect. The genuine result is a classification theorem for the MGD solution space itself -- that a single, physically sourced decoupler generates two mutually exclusive Lorentzian completions, with the metric selecting between them -- and, as noted in Sec.~\ref{sec4}, this classification depends only on the general hypotheses of Lemma~1, not on the specific analytic profile used to illustrate it in Sec.~\ref{sec6}.}

{This clarifies that the mechanism identified here does not reduce to counting zeros of $F(r;k)$: it requires specifying, for each $k$, which global completion is admissible, and we have shown that this is itself fixed by the requirement of a smooth Lorentzian spatial slice, not chosen by hand. We have further verified, for the explicit analytic example of Sec.~\ref{sec6}, that the resulting throat violates the null energy condition throughout the exterior region, consistent with the Morris--Thorne theorem \cite{Morris1988,Morris1988A}.}

{We reiterate, finally, that no change in spacetime topology is claimed anywhere in this paper, and that the classical no-go results of Geroch and Tipler \cite{Geroch1967,Tipler1977} -- which concern a single dynamical spacetime whose spatial topology changes in time -- accordingly do not apply to, and are not challenged by, the present construction: what we have established is that two static, non-diffeomorphic completions exist for the same decoupler $h(r)$, with Lemma~1 fixing which one is realized for a given value of the coupling $k$.}

{More broadly, this result suggests that the global properties of static spacetimes generated through gravitational decoupling are not determined solely by the local field equations of the $\theta$-sector: they also depend on which Lorentzian completion the resulting metric admits. The local matter-sector calculation fixes $h(r)$; it is the global requirement of Lorentzian smoothness, encoded in Lemma~1, that then decides whether the corresponding solution is read as a black hole or as a wormhole.}

Several directions for future work naturally emerge from this framework: extending the signature-obstruction argument to seeds beyond Schwarzschild, including charged and rotating configurations; exploring deformation sectors producing genuine bifurcation phenomena with pairs of roots; and quantifying lensing and time-delay signatures of the resulting geodesic structure.

\section*{ACKNOWLEDGMENTS}

Y. G\'omez-Leyton acknowledges to ANID subvenci\'on en la academia convocatoria a\~no 2025 85250186.

\bibliography{biblio.bib}
\bibliographystyle{elsarticle-num}

\end{document}